\documentclass[conference]{IEEEtran}
\usepackage[T1]{fontenc}
\usepackage{cite}
\usepackage{balance}
\usepackage{mathptmx,amsmath,amssymb,amsthm,booktabs,multirow,graphicx,microtype,url}
\usepackage[hidelinks,pdftitle={Spectrally Optimized Subinterval Trace Codes},pdflang=en]{hyperref}
\usepackage{algorithmic}
\usepackage{graphicx}
\usepackage{textcomp}
\usepackage{xcolor}
\newcommand{\F}{\mathbb F}
\newcommand{\C}{\mathcal{C}}
\newcommand{\J}{\mathcal{J}}
\newcommand{\B}{\mathcal{B}}
\newcommand{\T}{\mathcal{T}}
\newcommand{\Lch}{\mathcal{L}}
\newcommand{\rank}{\operatorname{rank}}
\newcommand{\argmin}{\operatorname*{arg\,min}}

\newcommand{\wt}{\operatorname{wt}}

\newtheorem{proposition}{Proposition}

\newtheorem{lemma}{Lemma}
\newtheorem{theorem}{Theorem}
\newtheorem{corollary}{Corollary}

\def\BibTeX{{\rm B\kern-.05em{\sc i\kern-.025em b}\kern-.08em
		T\kern-.1667em\lower.7ex\hbox{E}\kern-.125emX}}
\begin{document}
	
	\title{Affine Syndrome-Space Chase Decoding}
	
	\author{\IEEEauthorblockN{Majid Nasiri Khormuji}
		\IEEEauthorblockA{\textit{} \\
			\textit{1Finity Inc.}\\
			Kawasaki, Japan \\
			majid.n.k@ieee.org}
		\and
		\IEEEauthorblockN{Reza Dianat}
		\IEEEauthorblockA{\textit{Department of Electrical Engineering} \\
			\textit{Persian Gulf University}\\
			Bushehr, Iran \\
			dianat@pgu.ac.ir}
	}
	
	\maketitle
	
	\begin{abstract}
Chase decoding approaches maximum-likelihood performance by enlarging a reliability-generated list, but conventional realizations invoke an algebraic bounded-distance decoder for every test word. This letter realizes the same search in affine syndrome space. From the hard-decision syndrome, each test-word syndrome is obtained via a lightweight affine update over unreliable-coordinate parity-check columns. A tabulated syndrome-to-correction map replaces repeated decoder calls. The decoder preserves the Chase candidate set and metric decision, thereby enabling larger Chase lists without repeated algebraic decoder invocations.
\end{abstract}


\section{Introduction}
\IEEEPARstart{C}{hase} decoding remains an attractive bridge between hard-decision algebraic decoding and maximum-likelihood (ML) decoding. By perturbing a detected hard-decision word at unreliable coordinates, applying bounded-distance decoding (BDD) to the resulting test words, and selecting the best surviving candidate according to the channel metric, Chase decoding can approach ML performance with a sufficiently large reliability-generated list \cite{Chase1972}. The price is repeated algebraic decoding, whose cost grows with the list size and can dominate the overall complexity.

The same Chase search can be viewed from the syndrome domain. For a linear block code, the syndrome of every Chase test word is an affine update of the hard-decision syndrome, generated by the parity-check columns corresponding to the perturbed unreliable positions. Thus, the reliability-domain search over test words can also be realized as a \emph{syndrome-space search}. When the bounded-distance \emph{syndrome-to-correction map} is tabulated, repeated BDD calls can be replaced by affine syndrome updates, map evaluations, and metric comparisons, without changing the Chase candidate set or final decision.

For Reed--Solomon (RS), Bose--Chaudhuri--Hocquenghem (BCH), and related algebraic codes \cite{WickerBhargava1994}, this bottleneck has motivated fast and low-complexity Chase realizations. Stochastic and low-latency Chase variants have also been proposed for implementation-aware RS decoding \cite{Leroux2010,Jeong2021}. Wu's one-pass Chase decoder reduces repeated RS decoding work through locator-polynomial, interpolation, linear-feedback shift-register (LFSR), and transform-domain updates \cite{Wu2012}. Syndrome-based fast Chase decoding has also been developed using module-minimization and Gr\"obner-basis formulations for RS codes \cite{ShanyBerman2022}, and later extended to binary BCH codes through Wu-list decoding ideas \cite{ShanyBermanBCH2023}. Recent low-complexity Chase decoders further exploit module-based and channel-evaluation structures to reduce repeated RS decoding work \cite{XingModule2020,Wang2022}. Low complexity Chase decoder implementation for binary image of RS (255,239) is considered in \cite{Torres}. These works demonstrate the importance of reducing repeated algebraic work, but they rely on nontrivial polynomial, interpolation, module-update, or list-decoding machinery.

We focus on a complementary small-redundancy regime. Large Chase lists are attractive because they can move performance close to ML decoding, but conventional Chase becomes costly because each additional test word may require another BDD attempt. At the same time, the bounded-distance syndrome-to-correction map is small enough to store as a tabulated map. This creates a useful memory--computation tradeoff: repeated algebraic decoding can be replaced by syndrome addressing.

The proposed decoder is an \emph{exact affine syndrome-space realization} of two-choice Chase decoding. The reliability-domain search over test words is recast as a syndrome-space search, where all test-word syndromes are affine updates of the hard-decision syndrome. When the syndrome-to-correction map is tabulated, each trial reduces to an affine syndrome update, a map evaluation, and a metric comparison. The candidate set and final metric decision are identical to those of conventional Chase decoding.

We prove exact equivalence, establish the unambiguity of the syndrome-to-correction map within the decoding radius, and characterize the affine search dimension and complexity. Reed--Solomon simulations with $16$- and $64$-QAM verify identical decisions and show that larger lists approach ML performance without repeated decoder invocations.

\section{System Model and Chase Decoding}

Let \(\C\subseteq \F_q^n\) be an \([n,k]\) linear block code over
the finite field \(\F_q\) of size \(q\), with rate \(R=k/n\) and
parity-check matrix
\begin{equation}
    \mathbf{H}=[\mathbf{h}_1,\ldots,\mathbf{h}_n]\in\F_q^{r\times n},
    \label{eq:Hdef}
\end{equation}
where \(r=n-k\) is the redundancy and \(\mathbf{h}_j\) denotes the \(j\)-th column of \(\mathbf H\). A codeword of length \(n\) is denoted by
\begin{equation}
    \mathbf c=(c_1,\ldots,c_n)\in\C .
    \label{eq:cdef}
\end{equation}
Each field symbol is mapped to a modulation symbol through a one-to-one
labeling map
\begin{equation}
    \mu:\F_q\rightarrow\mathcal A,
    \label{eq:mudef}
\end{equation}
where \(\mathcal A=\{\mu(a):a\in\F_q\}\) is the signal constellation. The transmitted modulation vector is
\begin{equation}
    \mathbf x=(\mu(c_1),\ldots,\mu(c_n)).
    \label{eq:xdef}
\end{equation}
The channel output is \(\mathbf y=(y_1,\ldots,y_n)\). The channel is memoryless with transition law
\begin{equation}
    p(\mathbf y|\mathbf c)=\prod_{j=1}^{n}p(y_j|\mu(c_j)).
    \label{eq:memoryless}
\end{equation}
For complex additive white Gaussian noise (AWGN), let
\(\mathbf w=(w_1,\ldots,w_n)\) have independent entries distributed as
\(\mathcal{CN}(0,N_0)\), where \(N_0\) is the noise variance. The channel
output is
\begin{equation}
    y_j=\mu(c_j)+w_j.
    \label{eq:awgn}
\end{equation}
For each coordinate \(j\) and symbol \(a\in\F_q\), define the local negative log-likelihood metric
\begin{equation}
    \lambda_j(a)=-\log p(y_j|\mu(a)).
    \label{eq:lambda}
\end{equation}
For AWGN, the same ordering is obtained from the squared-distance metric
\begin{equation}
    \lambda_j(a)=|y_j-\mu(a)|^2,
    \label{eq:awgnmetric}
\end{equation}
where the common scaling by \(N_0\) and additive terms independent of \(a\) are omitted. Hence, the symbol-wise decisions and the final metric comparison are unchanged.

The detected hard-decision symbol at coordinate \(j\) is \(z_j\), and
the hard-decision word is \(\mathbf{z}_0\), where
\begin{equation}
    z_j=\argmin_{a\in\F_q}\lambda_j(a),
    \qquad
    \mathbf{z}_0=(z_1,\ldots,z_n)\in\F_q^n.
    \label{eq:harddecision}
\end{equation}
Ties in all symbol-wise minimizations are broken by a fixed deterministic rule. The hard-decision syndrome is
\begin{equation}
   \mathbf S_0=\mathbf{H}\mathbf{z}_0^T\in\F_q^r.
    \label{eq:S0def}
\end{equation}
Let \(a_{j,1}\) and \(a_{j,2}\) denote the most likely and second most likely symbols at coordinate \(j\), respectively, i.e.,
\begin{equation}
    a_{j,1}=z_j,
    \qquad
    a_{j,2}=\argmin_{a\in\F_q\setminus\{a_{j,1}\}}\lambda_j(a).
    \label{eq:bestsecond}
\end{equation}
The reliability gap is
\begin{equation}
    \rho_j=\lambda_j(a_{j,2})-\lambda_j(a_{j,1}).
    \label{eq:rhodef}
\end{equation}
Let
\begin{equation}
    \J=\{j_1,\ldots,j_L\}
    \label{eq:Jdef}
\end{equation}
be the \(L\) coordinates with the smallest reliability gaps. For each selected coordinate \(j_\ell\), define the symbol increment
\begin{equation}
    \Delta_\ell=a_{j_\ell,2}-a_{j_\ell,1}\in\F_q.
    \label{eq:Deltadef}
\end{equation}

A two-choice Chase pattern is
\begin{equation}
    \mathbf b=(b_1,\ldots,b_L)\in\B\subseteq\{0,1\}^L,
    \label{eq:bdef}
\end{equation}
where \(\B\) is the Chase pattern set, \(b_\ell=0\) keeps the hard-decision symbol at coordinate \(j_\ell\), and \(b_\ell=1\) replaces it by the second most likely symbol. The corresponding test word is
\begin{equation}
    \mathbf z(\mathbf b)
    =
    \mathbf{z}_0+
    \sum_{\ell=1}^{L}b_\ell\Delta_\ell\mathbf e_{j_\ell},
    \label{eq:testwordvector}
\end{equation}
where \(\mathbf e_j\) is the \(j\)-th unit vector. Thus, only selected coordinates with \(b_\ell=1\) are changed. The full Chase-\(2^L\) list corresponds to
\begin{equation}
    \B=\{0,1\}^L,
    \qquad
    B=|\B|=2^L .
    \label{eq:Bdef}
\end{equation}

Each test word is passed to a bounded-distance decoder (BDD). Equivalently, let
\begin{equation}
    \T:\F_q^r\rightarrow \F_q^n\cup\{\bot\}
    \label{eq:Tdef}
\end{equation}
be the BDD syndrome-to-correction map, where \(\bot\) is a failure symbol not contained in \(\F_q^n\). If
\begin{equation}
    \T\!\left(\mathbf H\mathbf z(\mathbf b)^T\right)=\mathbf e(\mathbf b)\neq\bot,
    \label{eq:BDDsuccess}
\end{equation}
then conventional Chase forms the candidate codeword
\begin{equation}
    \mathbf c(\mathbf b)=\mathbf z(\mathbf b)-\mathbf e(\mathbf b).
    \label{eq:candidate}
\end{equation}
The surviving candidates define the Chase candidate set
\begin{equation}
    \Lch_{\rm Ch}
    =
    \{\mathbf c(\mathbf b):\mathbf b\in\B,\
    \T(\mathbf H\mathbf z(\mathbf b)^T)\neq\bot\}.
    \label{eq:chasecandset}
\end{equation}
For any word \(\mathbf v=(v_1,\ldots,v_n)\in\F_q^n\), define the additive word metric
\begin{equation}
    \Lambda(\mathbf v)=\sum_{j=1}^{n}\lambda_j(v_j).
    \label{eq:wordmetric}
\end{equation}
The Chase decision is
\begin{equation}
    \hat{\mathbf c}=
    \argmin_{\mathbf c\in\Lch_{\rm Ch}}
    \Lambda(\mathbf c),
    \label{eq:metricdecision}
\end{equation}
which reduces to minimum Euclidean-distance selection under AWGN. The affine syndrome-space construction below preserves \(\Lch_{\rm Ch}\) and \eqref{eq:metricdecision} exactly; only the generation of test-word syndromes is changed.

\section{Affine Syndrome-Space Realization}

We now recast the Chase test-word search as a \emph{syndrome-space search}. The Chase candidate set in \(\Lch_{\rm Ch}\) is determined by the syndromes of the test words; once the hard-decision syndrome is known, each test-word syndrome is an affine update generated by selected parity-check columns. This affine representation follows from syndrome linearity, but its role here is algorithmic: the same two-choice Chase search can be realized by affine syndrome updates, evaluation of a syndrome-to-correction map, and metric comparison. When the map is tabulated, the evaluation becomes a lookup. Throughout this section, $\T$ denotes the bounded-distance decoder's syndrome-to-correction map introduced above, and all algebraic operations are over $\F_q$ unless stated otherwise.

\subsection{Affine Syndrome Representation}

For the unreliable coordinate $j_\ell$, define the syndrome increment
\begin{equation}
    \mathbf{g}_\ell=\Delta_\ell \mathbf{h}_{j_\ell}\in\F_q^r,
    \qquad \ell=1,\ldots,L .
    \label{eq:gdef}
\end{equation}
Collect these increments in the matrix
\begin{equation}
    \mathbf{G}_L=[\mathbf{g}_1,\ldots,\mathbf{g}_L]\in\F_q^{r\times L}.
    \label{eq:GLdef}
\end{equation}
The syndrome of the Chase test word in \eqref{eq:testwordvector} is
\begin{align}
    \mathbf S(\mathbf b)
    &=\mathbf{H}\mathbf z(\mathbf b)^T                                      \label{eq:syndrome1}\\
    &=\mathbf{H}\mathbf z_0^T+
      \sum_{\ell=1}^{L} b_\ell\Delta_\ell \mathbf{H}\mathbf e_{j_\ell}^T     \label{eq:syndrome2}\\
    &=\mathbf S_0+\sum_{\ell=1}^{L} b_\ell \mathbf g_\ell                           \label{eq:syndrome3}\\
    &=\mathbf S_0+\mathbf{G}_L\mathbf b .                                             \label{eq:affineidentity}
\end{align}
Thus, all Chase-pattern syndromes lie in the affine set
\begin{equation}
    \mathbf S_0+\operatorname{col}(\mathbf G_L),
    \label{eq:affinespace}
\end{equation}
where $\operatorname{col}(\mathbf G_L)$ denotes the column space of $\mathbf G_L$. The proposed implementation computes $\mathbf S_0$ once, forms $\mathbf g_1,\ldots,\mathbf g_L$, and evaluates each Chase pattern through \eqref{eq:affineidentity}, rather than recomputing $\mathbf H\mathbf z(\mathbf b)^T$ from scratch.

If the bounded-distance map returns
\begin{equation}
    \T(\mathbf S(\mathbf b))=\mathbf e(\mathbf b)\ne\bot,
    \label{eq:afftablequery}
\end{equation}
then the corresponding candidate is
\begin{equation}
    \mathbf c(\mathbf b)=\mathbf z(\mathbf b)-\mathbf e(\mathbf b),
    \label{eq:affcandidate}
\end{equation}
and the final decision is made using the same channel metric as in conventional Chase decoding.

\subsection{Syndrome-Map Injectivity and Exact Equivalence}

The following lemma shows that the syndrome-to-correction map is unambiguous inside the bounded-distance decoding radius.

\begin{lemma}[Injectivity of the syndrome-to-correction map]
Let $\mathcal C$ have minimum distance $d$, and let
\begin{equation}
    t=\left\lfloor\frac{d-1}{2}\right\rfloor .
    \label{eq:tdef}
\end{equation}
The syndrome map
\begin{equation}
    \mathbf e\mapsto \mathbf H\mathbf e^T
    \label{eq:syndromemap}
\end{equation}
is injective over
\begin{equation}
    \mathcal E_t=\{\mathbf e\in\F_q^n:\wt(\mathbf e)\le t\}.
    \label{eq:Etdef}
\end{equation}
\end{lemma}

\emph{Proof:} Assume that $\mathbf e'_1,\mathbf e'_2\in\mathcal E_t$ have the same syndrome. Then
\begin{equation}
    \mathbf H(\mathbf e'_1-\mathbf e'_2)^T=0,
    \label{eq:injectivekernel}
\end{equation}
so $\mathbf e'_1-\mathbf e'_2$ is a codeword of $\mathcal C$. Moreover,
\begin{equation}
    \wt(\mathbf e'_1-\mathbf e'_2)
    \le \wt(\mathbf e'_1)+\wt(\mathbf e'_2)
    \le 2t<d .
    \label{eq:injectiveweight}
\end{equation}
By definition of the minimum distance, the only codeword with weight smaller than $d$ is the zero codeword. Hence $\mathbf e'_1-\mathbf e'_2=\mathbf 0$, and therefore $\mathbf e'_1=\mathbf e'_2$. \hfill$\square$

Consequently, the number of nonfailure entries in a tabulated realization of the map is
\begin{equation}
    N_\T=\sum_{w=0}^{t}\binom{n}{w}(q-1)^w,
    \label{eq:tablesize}
\end{equation}
although the full syndrome address space has size $q^r$.

\begin{theorem}[Exact equivalence]
For any fixed Chase pattern set $\B$ and the same BDD correction map $\T$, affine syndrome-space Chase decoding produces the same surviving candidate set and the same final metric decision as conventional Chase decoding.
\end{theorem}

\emph{Proof:} For a pattern $\mathbf b\in\B$, conventional Chase computes the syndrome $\mathbf H\mathbf z(\mathbf b)^T$. By \eqref{eq:affineidentity}, affine syndrome-space Chase computes exactly the same syndrome,
\begin{equation}
    \mathbf H\mathbf z(\mathbf b)^T=\mathbf S_0+\mathbf G_L\mathbf b=\mathbf S(\mathbf b).
    \label{eq:samesyndrome}
\end{equation}
Thus both decoders query the same map $\T$ with the same argument. If $\T$ returns $\bot$, both decoders reject the pattern. If $\T$ returns $\mathbf e(\mathbf b)$, both decoders form the same candidate codeword in \eqref{eq:affcandidate}. Therefore the surviving candidate sets are identical. Since both decoders apply the same metric rule in \eqref{eq:metricdecision}, the final decision is identical. \hfill$\square$

\noindent\textbf{Affine syndrome-space implementation.} The decoder can therefore be implemented as follows. First compute the hard-decision syndrome $\mathbf S_0$ and the increments $\mathbf g_\ell=\Delta_\ell \mathbf h_{j_\ell}$. For each pattern $\mathbf b\in\B$, form $\mathbf S(\mathbf b)$ by \eqref{eq:affineidentity}, evaluate $\T$ (or look it up when tabulated), form the surviving candidate in \eqref{eq:affcandidate}, and finally apply the metric rule \eqref{eq:metricdecision}. This implementation changes neither the Chase pattern set nor the final decision rule; it only changes how the BDD inputs are obtained.

\subsection{Affine Dimension, Collisions, and MDS Codes}

The affine representation also describes the dimension and possible
collisions of the Chase syndrome set.

\begin{proposition}[Affine dimension and collisions]
Let
\begin{equation}
    \rho_L=\rank(\mathbf G_L).
    \label{eq:rho}
\end{equation}
Then
\begin{equation}
    \rho_L\le \min\{L,r\}.
    \label{eq:rhobound}
\end{equation}
Moreover, the number of distinct Chase syndromes satisfies
\begin{equation}
    |\{\mathbf S(\mathbf b):\mathbf b\in\B\}|
    \le \min\{|\B|,q^{\rho_L}\},
    \label{eq:lookupbound}
\end{equation}
and two patterns collide in syndrome space if and only if
\begin{equation}
    \mathbf S(\mathbf b)=\mathbf S(\mathbf b')
    \quad\Longleftrightarrow\quad
    \mathbf G_L(\mathbf b-\mathbf b')=0 .
    \label{eq:collision}
\end{equation}
\end{proposition}

\emph{Proof:} Since \(\mathbf G_L\in\F_q^{r\times L}\), its rank cannot exceed
its number of columns or rows, which proves \eqref{eq:rhobound}. The
syndrome set is contained in the affine space \(\mathbf S_0+\operatorname{col}(\mathbf G_L)\), which has \(q^{\rho_L}\) elements, and it is generated by only
\(|\B|\) queried patterns. This proves \eqref{eq:lookupbound}. The
collision condition follows by substituting \(\mathbf S(\mathbf b)=\mathbf S_0+\mathbf G_L\mathbf b\)
and cancelling \(\mathbf S_0\). \hfill$\square$

For two-choice Chase sets over nonbinary fields, the exact collision count
need not be uniform; however, \eqref{eq:collision} still characterizes
all collisions and \eqref{eq:lookupbound} remains valid.

\begin{corollary}[Rank saturation for maximum-distance separable codes]
For a maximum-distance separable (MDS) code, including the RS$(n,k)$ codes considered here, assume the unreliable coordinates $j_1,\ldots,j_L$ are distinct and $\Delta_\ell\ne0$ for all $\ell$. Then
\begin{equation}
    \rho_L=\min\{L,r\}.
    \label{eq:rsrank}
\end{equation}
\end{corollary}

\emph{Proof:} An $[n,k]$ MDS code has a parity-check matrix in which any $r=n-k$ columns are linearly independent. Multiplying a selected column $\mathbf h_{j_\ell}$ by the nonzero scalar $\Delta_\ell$ does not change linear dependence. Hence any subset of at most $r$ columns of $G_L$ is independent. If $L<r$, the rank is $L$; if $L\ge r$, the rank cannot exceed $r$ and reaches $r$ by the same independence property. This proves \eqref{eq:rsrank}. \hfill$\square$

The results above show that the proposed construction changes implementation rather than decoding behavior. The next section quantifies the resulting cost reduction.

\begin{figure*}[!t]
\centering
\begin{minipage}[t]{0.5\textwidth}
\centering
\vspace{-3mm}
\includegraphics[width=1.0\linewidth]{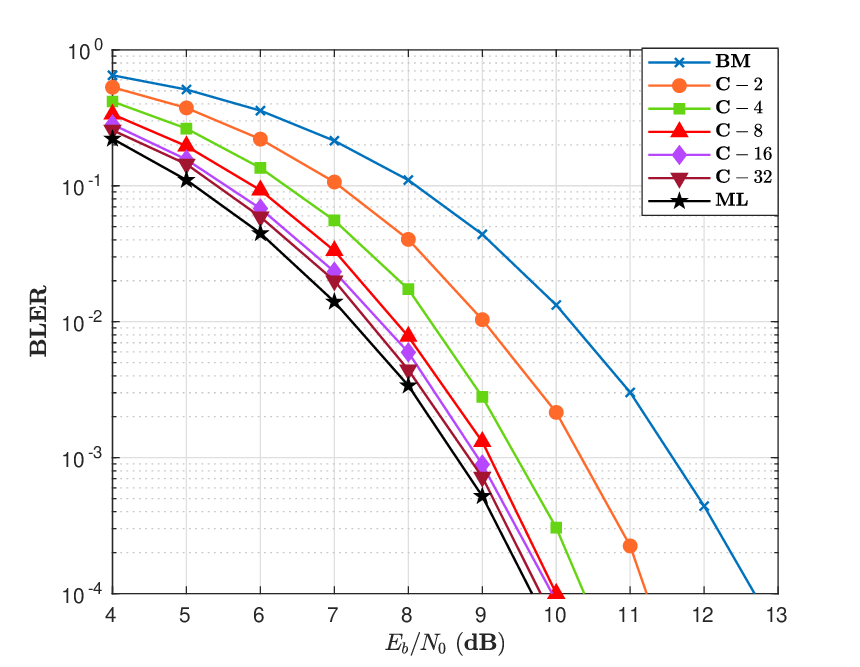}
\vspace{-8mm}
\caption{BLER of RS$(6,4)$ over $\F_{16}$ with Gray-labeled $16$-QAM.}
\label{fig:rs64bler}
\end{minipage}\hfill
\begin{minipage}[t]{0.5\textwidth}
\centering
\vspace{-3mm}
\includegraphics[width=1.0\linewidth]{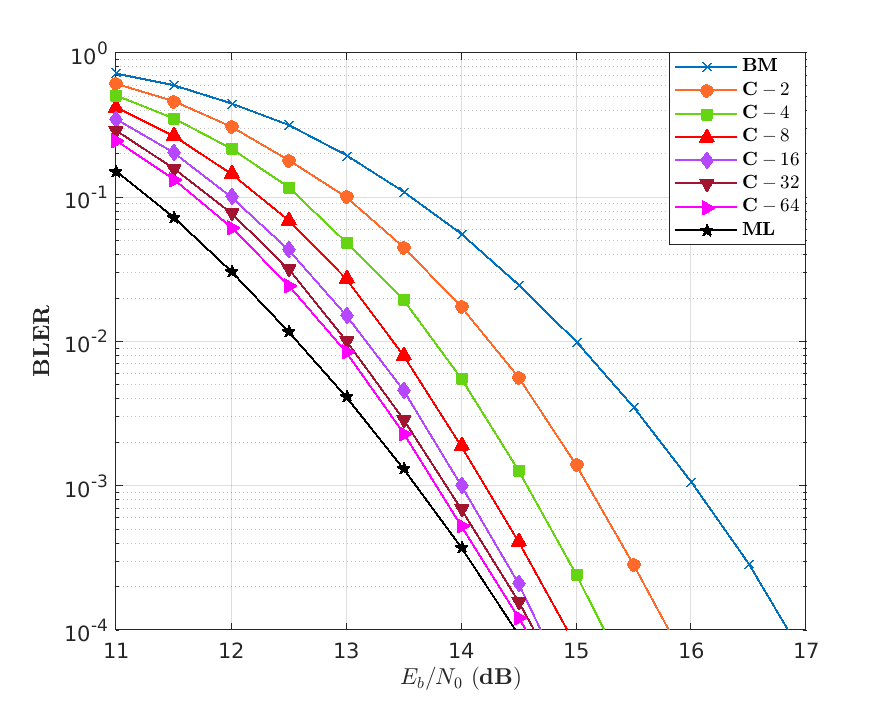}
\vspace{-8mm}
\caption{BLER of RS$(20,18)$ over $\F_{64}$ with Gray-labeled $64$-QAM.}
\label{fig:rs2018bler}
\end{minipage}
\end{figure*}

\section{Realization Complexity and Memory}

The equivalence theorem separates decoding performance from realization
cost. Since affine syndrome-space Chase preserves the Chase candidate set
and metric decision, its value is measured by the cost of realizing the
same pattern set. We compare direct Chase, affine syndrome-space Chase with
independent syndrome updates, and Gray-code affine syndrome-space Chase.
All costs below are normalized operation counts; metric scoring is common
to all realizations and is omitted.

Let $B=|\B|$ be the number of Chase patterns, and let $C_{\rm BDD}$
denote the cost of one algebraic bounded-distance decoding operation once
the syndrome is available. A direct Chase realization forms one syndrome
and invokes one bounded-distance decoder for every test word. Its algebraic
cost is modeled as
\begin{equation}
    C_{\rm dir}=B(nr+C_{\rm BDD}),
    \label{eq:directcost}
\end{equation}
where $nr$ accounts for multiplying the $r\times n$ parity-check matrix
by a length-$n$ test word.

Affine syndrome-space Chase forms the hard-decision syndrome once and
precomputes the $L$ column increments $\mathbf g_1,\ldots,\mathbf g_L$ at one-time cost
$Lr$. If a pattern has weight $w$, its syndrome is obtained by adding $w$
precomputed length-$r$ column increments. Since this precomputation is
one-time and independent of the list traversal, it is omitted from the
normalized curves below. For a general pattern set, define
\begin{equation}
    \bar w=\frac{1}{B}\sum_{\mathbf b\in\B}\wt(\mathbf b).
    \label{eq:avgweight}
\end{equation}
With one normalized map lookup per pattern, the affine cost is
\begin{equation}
    C_{\rm aff}=nr+B\bar w r+B .
    \label{eq:affcostgeneral}
\end{equation}
For the full two-choice list $\B=\{0,1\}^L$, each selected coordinate is
active in half of the patterns, hence
\begin{equation}
    \bar w=\frac{L}{2},
    \label{eq:avgweightfull}
\end{equation}
and
\begin{equation}
    C_{\rm aff}=nr+\frac{BL}{2}r+B .
    \label{eq:affcost}
\end{equation}

A further reduction follows by traversing the full two-choice list in
binary Gray-code order. Consecutive patterns differ in one component, so
if $\mathbf b_m$ and $\mathbf b_{m+1}$ differ only at index $\ell_m$, then
\begin{equation}
    \mathbf S(\mathbf b_{m+1})=\mathbf S(\mathbf b_m)\pm \mathbf g_{\ell_m}.
    \label{eq:grayupdate}
\end{equation}
The sign depends on the transition direction and is immaterial in
characteristic two. After computing $\mathbf S_0$, each additional pattern requires
one column-syndrome update and one map lookup. The Gray-code affine cost is
\begin{equation}
    C_{\rm aff}^{\rm Gray}=nr+(B-1)r+B .
    \label{eq:graycost}
\end{equation}
Gray-code ordering only changes the traversal of the same pattern set; it
does not change the candidate set, metric rule, or Chase decision.

The corresponding algebraic operation-count reduction is
\begin{equation}
    \Gamma_{\rm Gray}
    =
    \frac{C_{\rm dir}}{C_{\rm aff}^{\rm Gray}}
    =
    \frac{B(nr+C_{\rm BDD})}
    {nr+(B-1)r+B}.
    \label{eq:graygain}
\end{equation}
For fixed redundancy $r$, the reduction grows with both the block length
and the Chase list size: direct Chase repeats the $nr$-scale syndrome work
for every test word, whereas the affine realization forms the hard-decision
syndrome once and then updates in dimension $r$.

The memory cost is governed by the tabulated syndrome-to-correction map in
\eqref{eq:tablesize}. In the high-rate case $t=1$, this becomes
\begin{equation}
    N_{\T}=1+n(q-1),
    \label{eq:t1tablesize}
\end{equation}
which grows linearly in $n(q-1)$. This is the regime where tabulation is
most attractive: the map is modest, while direct Chase still repeats
syndrome formation and bounded-distance decoding across many test words.

\section{Numerical Results}
\label{sec:numerical}

We evaluate the Chase lists on two Reed--Solomon (RS) coded-transmission benchmarks \cite{ReedSolomon1960,WickerBhargava1994}. The first uses RS$(6,4)$ over $\F_{16}$ with Gray-labeled $16$-QAM, for which exhaustive codebook maximum-likelihood (ML) decoding is feasible. The second uses RS$(20,18)$ over $\F_{64}$ with Gray-labeled $64$-QAM, for which the ML reference is obtained by Wolf's syndrome-trellis decoder \cite{Wolf1978}. This codes has been evaluated with other decoders \cite{qgrand}, but here the focus is on the pattern-list size effect on the performance of Chase decoding. The channel model, hard decision, and metric are those of Section~II. All constellations are normalized to unit average symbol energy, and
\begin{equation}
    E_s/N_0=R\log_2(q)\,E_b/N_0,
    \label{eq:snrconversion}
\end{equation}
where $R=k/n$.

The BLER figures compare Berlekamp--Massey (BM) hard decoding \cite{Berlekamp1968,Massey1969}, Chase decoding with several values of $B$, and the corresponding ML reference, where $B=|\B|$ is the number of test patterns. The legend entry C-$B$ denotes Chase with $B$ test patterns, e.g., C-$32$ and C-$64$. By Theorem~1, the proposed realization has exactly the same candidate set and metric decision as Chase; hence every Chase curve is also its block-error-rate (BLER) curve. We first quantify the performance value of enlarging $B$ and then evaluate the realization cost of the same lists.

Fig.~\ref{fig:rs64bler} shows that larger lists give a clear soft-decision gain over BM and steadily reduce the gap to exhaustive ML for RS$(6,4)$. At BLER $10^{-3}$, C-$32$ gains about $2.6$ dB over BM and leaves about $0.3$ dB to exhaustive ML. Fig.~\ref{fig:rs2018bler} shows the high-rate RS$(20,18)$ benchmark. The same trend appears with the Wolf-ML reference: at BLER $10^{-3}$, C-$64$ gains about $2.2$ dB over BM and remains about $0.2$ dB from Wolf ML. Thus, moderate Chase lists already close most of the gap to ML.

\begin{figure}[!t]
	\vspace{-8mm}
\centering
\includegraphics[width=0.98\columnwidth]{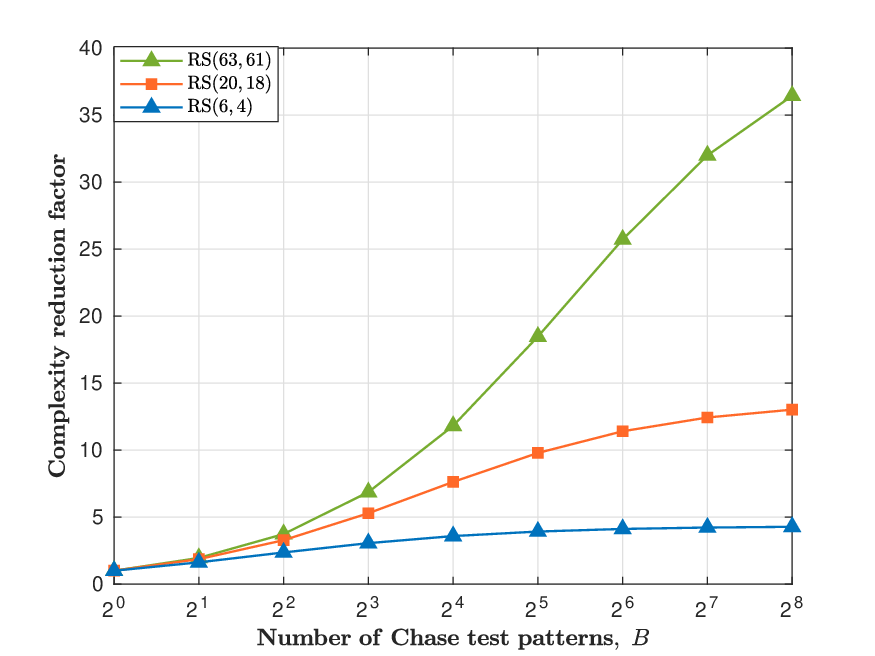}
\vspace{-3mm}
\caption{Operation-count reduction relative to direct Chase.}
\label{fig:complexity}
\end{figure}

Fig.~\ref{fig:complexity} then evaluates the operation-count reduction predicted by \eqref{eq:graygain} for the same lists. For the BLER operating points above, the Gray-code affine realization gives about $3.9\times$ reduction for RS$(6,4)$ at $B=32$ and $11.4\times$ for RS$(20,18)$ at $B=64$. The longer RS$(63,61)$ curve reaches about $25.7\times$ at $B=64$ and $36.5\times$ at $B=256$, illustrating the growth with block length when $r$ is fixed. The normalization $C_{\rm BDD}=1$ treats one algebraic BDD operation as only one map lookup; hence the plotted gains are conservative for any implementation in which a BDD call costs more than a tabulated-map lookup.

\section{Concluding Remarks}

This letter presented affine syndrome-space Chase decoding, an exact syndrome-domain realization of two-choice Chase decoding. The method preserves the Chase pattern set, candidate set, metric decision, and BLER, while replacing repeated algebraic decoding attempts by affine syndrome updates and tabulated syndrome-map lookups.

The results show the two sides of the tradeoff. Moderate lists close most of the BM-to-ML performance gap, while the Gray-code affine realization gives about $3.9\times$ and $11.4\times$ operation-count reduction for the RS$(6,4)$ and RS$(20,18)$ cases considered. Since the complexity model counts one algebraic BDD operation as only one map lookup, the practical savings can be larger whenever bounded-distance decoding is more costly than tabulated-map access. The method is therefore most attractive for small-redundancy algebraic codes, where the syndrome-to-correction map is compact and repeated BDD calls dominate.

\end{document}